\documentclass[preprint,amsmath,amssymb,aps,notitlepage]{revtex4-2}

\usepackage{graphicx}
\usepackage{dcolumn}
\usepackage{bm}
\usepackage[normalem]{ulem} 

\usepackage[spanish,english]{babel}
\usepackage[utf8]{inputenc} 
\usepackage{xcolor}

\begin{document}

\title{Termoelectricidad cuántica: conversión de energía a escala nanométrica}

\author{David Sánchez}
\affiliation{Instituto de Física Interdisciplinar y Sistemas Complejos IFISC (UIB-CSIC), Campus Universitat Illes Balears, 07122 Palma de Mallorca}%

\author{Rafael Sánchez}%
\affiliation{Departamento de Física Teórica de la Materia Condensada y Centro de Investigación de Física de la Materia Condensada (IFIMAC), Universidad Autónoma de Madrid, 28049 Madrid}

\date{\today}

\begin{abstract}

La termoelectricidad abarca un conjunto de fenómenos
por los cuales el calor puede transformarse en electricidad,
y viceversa. Las implicaciones prácticas de esta conversión son obvias, si bien el estudio del
efecto termoeléctrico también contribuye a comprender en profundidad del comportamiento de los electrones en materiales conductores. En este trabajo, se exponen los resultados más interesantes obtenidos recientemente
con conductores ultrapequeños, donde el comportamiento electrónico es puramente cuántico y las teorías clásicas de termoelectricidad deben, en consecuencia, revisarse a fondo.
\\
\indent Thermoelectricity describes phenomena that allow us to convert heat into electricity and vice versa. The practical implications of this conversion are
obvious. However, research on thermoelectric effects also leads to
a deep understanding of the electronic properties
in conductive materials. In this paper,
we discuss the most interesting results obtained
recently in ultrasmall conductors, where the electronic
behavior is purely quantum and classical theories
of thermoelectricity are thus to be revised in detail.
\end{abstract}

\maketitle

\section{Introducción}

\begin{figure}[t]
\includegraphics[width=0.5\linewidth]{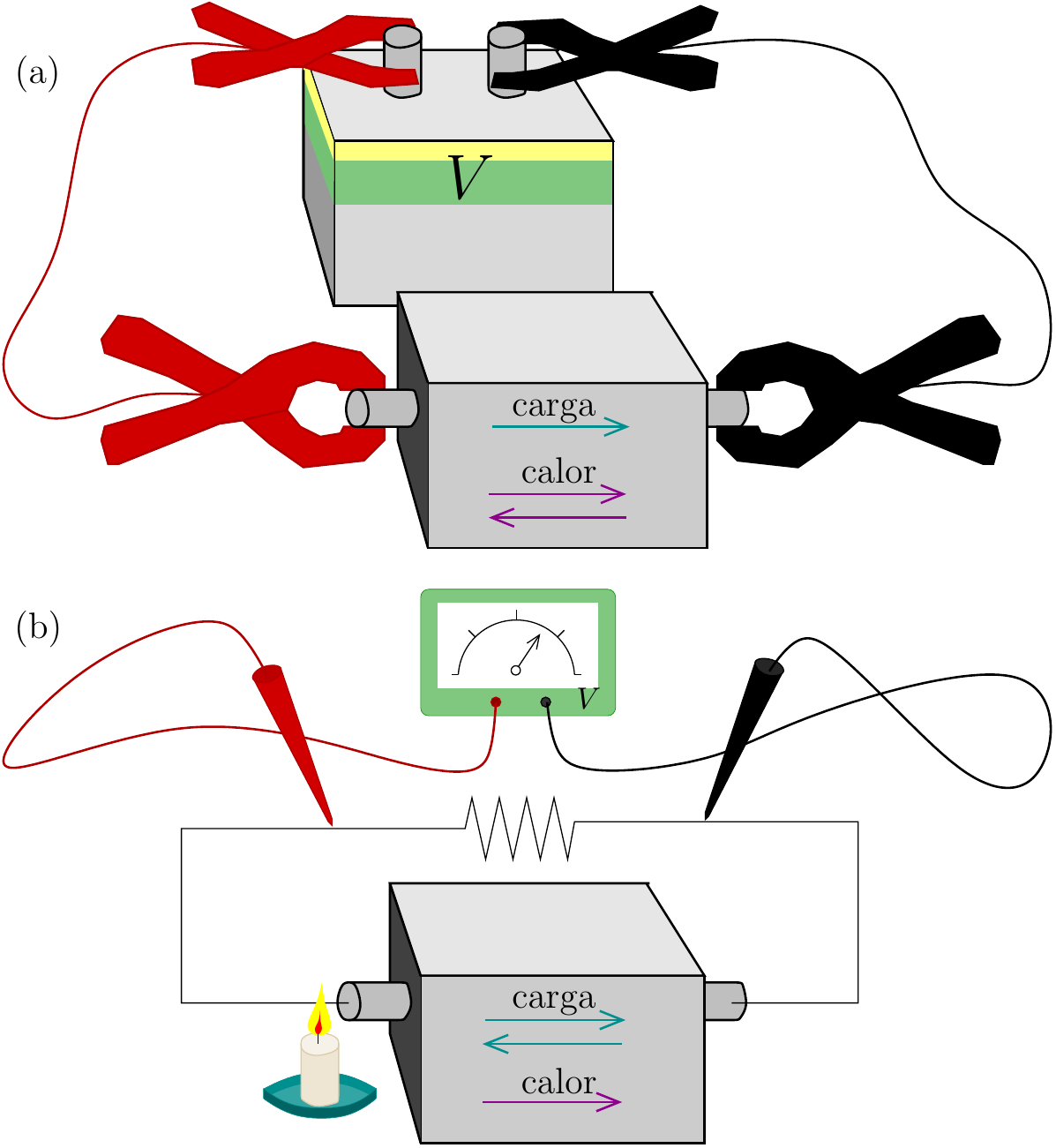}
\caption{\label{corrientes}
(a) La diferencia de potencial $V$ suministrada por una
batería propicia un flujo de carga a través del material conductor. Igualmente, $V$ puede generar un flujo de calor mediante el efecto Peltier. A diferencia de la carga,
el calor de Peltier puede fluir en cualquier dirección
(lo que dependerá de las propiedades del conductor).
(b) La carga también puede transportarse mediante una diferencia
de temperaturas aplicada entre los extremos del conductor
(efecto Seebeck).
Existe un paralelismo con el caso anterior: mientras que el sentido de la corriente de Seebeck no está fijo, el de la corriente de calor generada térmicamente viene determinado por el segundo principio de la termodinámica.}
\end{figure}

Las primeras décadas del s.~XIX fueron testigo de un florecimiento de las investigaciones sobre fenómenos eléctricos gracias a la invención de la batería, un dispositivo que permite generar corrientes eléctricas. Así, Georg Ohm demostró que la corriente $I$ (es decir, la carga por unidad de tiempo) que circula por un conductor es directamente proporcional a la diferencia de potencial $V$ existente entre los bornes de la batería (veáse la figura ~\ref{corrientes}(a)): $I=GV$, siendo la conductancia $G$ un coeficiente cinético. El estudio de $G$ abre la puerta a clasificar los materiales, atendiendo a la facilidad con que los portadores de carga fluyen, como conductores (grandes valores de $G$, como los metales) o como aislantes eléctricos (pequeños valores de $G$, como los plásticos o las cerámicas).

Sin embargo, la aplicación de un voltaje $V$ no es la única forma de inducir una corriente eléctrica. Un poco antes que Ohm, Thomas Seebeck descubrió que una diferencia de temperaturas $\Delta T$ podía generar un movimiento de cargas entre los extremos frío y caliente del conductor, como se muestra en la figura~\ref{corrientes}(b). Este hallazgo inauguró la termoelectricidad, un campo que resultó extraordinariamente fecundo no solo porque tendía un puente entre dos disciplinas que parecían hasta entonces poco relacionadas (la termodinámica y el electromagnetismo), sino porque ofrecía soluciones alternativas a la construcción de dispositivos básicos como los motores térmicos y los refrigeradores.

Agrupando las contribuciones eléctricas y termoeléctricas, la ley de Ohm se generaliza a $I=GV+L\Delta T$, donde $L$ es la respuesta termoeléctrica, otro coeficiente cinético que, al igual que G, depende de las propiedades del material conductor. El hecho de que coexistan dos fuerzas electromotrices ($V$ y $\Delta T$) posibilita la cancelación de la corriente $I$. En efecto, dada una diferencia de temperaturas, si imponemos $I=0$ en la ecuación anterior, hay un termovoltaje $V_t\equiv V(I=0)=-L \Delta T/G$, para el cual la corriente generada eléctricamente es igual, pero de signo opuesto, a la corriente inducida térmicamente. Al cociente $S\equiv V_t/\Delta T=-L/G$ se le denomina coeficiente de Seebeck o termopotencia. Como veremos más adelante, la termopotencia contiene información valiosa sobre las propiedades físicas del material sujeto a estudio.

La íntima conexión entre termodinámica y electromagnetismo se puso de manifiesto de forma aun más evidente en los trabajos posteriores de Jean Peltier y Lord Kelvin.
Años antes, Joseph Fourier había llegado a la conclusión de que, si se sometía un material a un gradiente térmico $\Delta T$, el calor transmitido por unidad de tiempo $J$ debía ser proporcional a $\Delta T$, de forma similar a lo que ocurría con la ley de Ohm para la conducción de carga: $J = K \Delta T$. En esta ecuación, $K$ se denomina conductancia térmica y, análogamente al caso eléctrico, se puede establecer una clasificación de los materiales en función de su capacidad de conducir calor: conductores y aislantes térmicos. Curiosamente, los metales son buenos conductores eléctricos y térmicos, lo que nos hace sospechar que es la misma partícula la responsable de transportar la carga y parte del calor. Pero detengámonos un momento en el descubrimiento de Peltier. Este comprobó que un voltaje $V$ podía poner en marcha un flujo de calor a lo largo del conductor y no únicamente una corriente eléctrica: $J = M V$. Este tipo de calor es de naturaleza diferente al que observó James Joule un poco más tarde, también cuando generó calor mediante corrientes eléctricas. Mientras que el calor de Peltier es reversible, es decir, puede utilizarse tanto para calentar como para enfriar, el de Joule es irreversible porque obedece a procesos disipativos que liberan energía y, por tanto, siempre calientan el material. El efecto Joule hace que un dispositivo se caliente al funcionar (pensemos en cualquier electrodoméstico), mientras que el de Peltier puede utilizarse para construir un refrigerador. La explicación de esta diferencia tan significativa la dio Lord Kelvin al advertir que el calor de Peltier era, en realidad, un efecto termoeléctrico más; de hecho, hay una relación insospechada e increíblemente simple entre los coeficientes de Seebeck y Peltier: $M = T L$, siendo $T$ la temperatura promedio del material. Hubo que esperar casi cien años hasta que el genio de Lars Onsager proporcionó la razón profunda que subyace a la universalidad de la igualdad de Kelvin: las relaciones de reciprocidad (véase el cuadro adjunto).

\fbox{\begin{minipage}{35em}\footnotesize
{\bf Relaciones de reciprocidad}\\
Imaginemos un sistema homogéneo en equilibrio. El segundo principio de la termodinámica establece que su entropía es máxima. ¿Qué pasa fuera del equilibrio? Dentro del formalismo de la termodinámica de procesos irreversibles~\cite{mazur}, se puede calcular la tasa de producción de entropía $\Sigma$ (entropía por unidad de tiempo) debida a la acción de las distintas fuerzas actuando sobre el sistema. Se halla que $\Sigma$ viene dada por los flujos que atraviesan el sistema llevando energía, carga o densidad multiplicados por los coeficientes cinéticos definidos anteriormente. En consecuencia, fuera del equilibrio la entropía puede transportarse de un sitio a otro, si bien, globalmente, ha de crecer. Esto implica que $\Sigma>0$ y de aquí se infiere que las conductancias eléctrica y térmica son siempre positivas: $G, K>0$. Esto es consistente con dos hechos bien conocidos: 1) la potencia disipada de Joule es $GV^2$, y por tanto siempre positiva; 2) el calor se transfiere de un contacto caliente a uno frío, y no al revés, de acuerdo con el segundo principio, por lo que, aplicando la ley de Fourier, se tiene que $K>0$. Esto explica las direcciones fijas de la figura~\ref{corrientes}. Ahora bien, los coeficientes $L$ y $M$ pueden tomar cualquier signo, aunque deben obedecer la relación de Kelvin $M = T L$. Onsager pudo  demostrarlo basándose en una relación de reciprocidad que se deduce del principio de reversibilidad de los procesos microscópicos. Este dicta que cualquier proceso microscópico y su reverso temporal ocurren, en promedio, con la misma probabilidad.
\end{minipage}}

En resumen, la existencia de principios físicos muy generales, como la positividad de la producción de entropía o la simetría temporal de las ecuaciones microscópicas, impone res\-tric\-ciones a la posible respuesta que pueden presentar los conductores cuando se encuentran sometidos a fuerzas eléctricas y térmicas. Hasta ahora, nuestra descripción del fenómeno ha sido estrictamente clásica. ¿Qué ocurrirá si nos sumergimos en el mundo cuántico? Esta pregunta la abordaremos en la próxima sección.

\section{Conductores cuánticos}

\begin{figure}[t]
\includegraphics[width=0.5\linewidth]{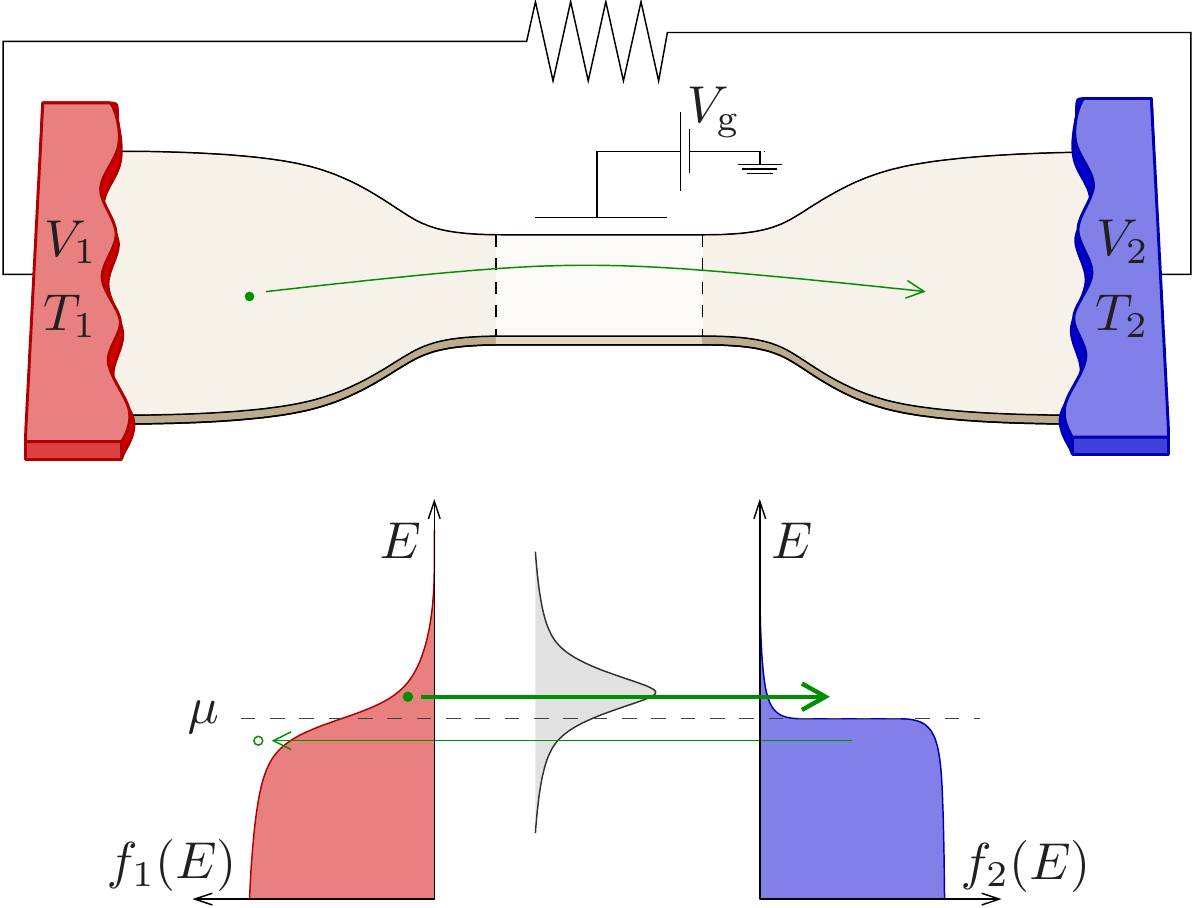}
\caption{\label{seebeck}
Ilustración esquemática de un conductor nanoscópico (delimitado por líneas discontinuas) conectado a dos terminales macroscópicos caracterizados, respectivamente, por potenciales $V_1$ y $V_2$
y temperaturas $T_1>T_2$. El diagrama
de energías se muestra debajo, donde la ocupación de las bandas de conducción de los contactos está representada por funciones de Fermi ($f_1(E)$ y $f_2(E)$ en función de la energía $E$) con el mismo potencial químico, $\mu$, y distinta temperatura. La nanoestructura actúa como un filtro de energías de tal forma que el flujo de electrones es máximo cuando el nivel de energía del conductor, cuya posición puede ajustarse
con un potencial de puerta $V_{\rm g}$, está en el rango energético de las excitaciones térmicas, alrededor de $\mu$.
Al incrementar la temperatura de un contacto, se crea
una corriente termoeléctrica que alimenta un circuito representado por la resistencia. En configuración de cortocircuito, el potencial químico de ambos terminales coincide, por lo que se genera una corriente, pero no potencia. Si el circuito está abierto, no hay corriente, pero se genera una diferencia de potencial $V_t$ o trabajo útil.}
\end{figure}

Se denominan conductores cuánticos o mesoscópicos a aquellos sistemas cuyo tamaño es lo suficientemente pequeño como para que los portadores de carga y energía retengan todas sus propiedades cuánticas mientras viajan por ellos. Estos sistemas se han construido en el laboratorio solo desde la década de 1980, aunque en la actualidad disponemos de una gran variedad de ellos~\cite{Ihn}. Por ejemplo, en la intercara entre ciertos semiconductores pueden formarse regiones de una pureza tal que los electrones fluyen a través de ellas balísticamente. Otros dispositivos útiles son las uniones moleculares, cuyo tamaño no excede 1-2~nm~\cite{cuevas}.
Finalmente, hoy en día también se pueden llevar a cabo experimentos de transporte termoeléctrico con nubes de átomos ultrafríos. Lo que tienen en común todos estos sistemas es la posibilidad de que el transporte de energía, carga o partículas manifieste señales de coherencia cuántica. En tal caso, al llevar el sistema fuera del equilibrio mediante campos externos, los portadores disipan su energía lejos del conductor, más precisamente en los contactos macroscópicos a los que está acoplado. 

En el dibujo superior de la figura~\ref{seebeck}, se muestra el diagrama de un conductor cuántico típico, conectado a sendos terminales a los que se les ha aplicado una diferencia de potencial y de temperatura.
Estos gradientes electrotérmicos permiten que los electrones atraviesen el conductor
tal y como indica la flecha verde.
En el gráfico inferior, se describe el conductor mediante un nivel energético discreto más un cierto ensanchamiento (la región sombreada) que proviene de la probabilidad que tiene un electrón de propagarse a cualquiera de los dos terminales. Los primeros experimentos con un sistema parecido se realizaron a fines del siglo pasado \cite{molenkamp}, mostrando un coeficiente de Seebeck $S$ con un perfil antisimétrico muy característico. Cuando se desplaza la posición del nivel cuántico mediante un potencial $V_{\rm g}$ aplicado a un terminal de puerta adyacente, el nivel cruza la energía de Fermi $\mu$ de los terminales. Esta energía es especial porque los electrones de conducción en el caso de fuerzas motrices pequeñas poseen un estado energético que se encuentra alrededor de $\mu$. El resto de los electrones no puede alcanzar el otro terminal, pues el principio de exclusión de Pauli se lo impide al estar los estados ya ocupados.
Como se puede ver en el panel inferior de la figura~\ref{seebeck}, los electrones que se encuentran situados por debajo de $\mu$ contribuyen a $S$ de forma positiva (flecha de trazo fino), mientras que los que están térmicamente excitados por encima de $\mu$ aportan un signo negativo a la termopotencia (flecha de trazo grueso). De esta forma, la respuesta termoeléctrica nos proporciona información sobre el carácter de los portadores (tipo electrón o hueco), que no puede extraerse con una medida puramente eléctrica. Es más: a bajas temperaturas, los coeficientes $S$ y $G$ están relacionadas a través de la ley de Mott: $S\propto (1/G) dG/dE$, donde la derivada se evalúa en $E=\mu$. Por tanto, una medida del coeficiente de Seebeck puede funcionar como complementaria a una de la conductancia. Merece la pena, además, indicar que mientras $G$ mide la densidad de estados del conductor, el coeficiente $S$ refleja la asimetría de la misma. Este resultado es importante por cuanto sistemas con simetría electrón-hueco, como los superconductores, no reaccionan termoeléctricamente ya que $S=0$. Las desviaciones de este punto pueden, por tanto, ofrecer datos sobre posibles rupturas de la simetría electrón-hueco en sistemas de baja dimensionalidad.

¿Se siguen obedeciendo las relaciones de reciprocidad en el régimen cuántico? Por un lado, sí se satisfacen, dado que el transporte cuántico, al menos para electrones independientes, puede describirse completamente con el formalismo de dispersión. Las matrices de dispersión se calculan a partir de las funciones de onda y, por consiguiente, representan la amplitud de probabilidad de que una partícula se transmita de un terminal a otro. En la medida que las funciones de onda cumplen la ecuación de Schrödinger, y esta es una ecuación simétrica bajo inversiones temporales, las relaciones de reciprocidad se siguen cumpliendo en conductores cuánticos. Por otro lado, el hecho de que la disipación ocurra lejos del conductor significa que los procesos de dispersión en el sistema mesoscópico son elásticos. Esto introduce unas nuevas relaciones de simetría estrictamente cuánticas, no contempladas en las relaciones originales de Onsager y que se han comprobado experimentalmente hace poco midiendo la respuesta termoeléctrica de un rectificador cuántico multiterminal~\cite{matthews}. Es interesante observar que, si se rompen estas relaciones cuánticas, esto querría decir que el transporte en una muestra dada deja de ser balístico y pasa a comportarse como en un conductor macroscópico, donde la longitud de termalización es mucho más pequeña que el tamaño del sólido. Así pues, las relaciones cuánticas termoeléctricas sirven para discernir el mecanismo concreto que difunde los portadores a través de un material. Lejos del equilibrio, tanto estas relaciones como la reciprocidad de Onsager dejan de cumplirse debido a que las fuerzas aplicadas son muy intensas y el sistema empieza a responder no linealmente a los gradientes externos. Los análisis del régimen no lineal de transporte son harto difíciles en sólidos macroscópicos, dadas las enormes distancias sobre las que se extienden los gradientes; en cambio, los conductores cuánticos nanoscópicos son excelentes bancos de pruebas para este propósito~\cite{david}. Por ejemplo, en una unión molecular se ha alcanzado un gradiente de temperatura de más de 1~GK/m \cite{kim}.

Aún más espectacular es la conexión entre la termopotencia y la entropía. Esta es una cantidad fundamental en física, pero es extraordinariamente difícil de medir, sobre todo en sistemas cuánticos. El coeficiente de Seebeck puede entonces ser de gran ayuda. La causa la tenemos que buscar, de nuevo, en las relaciones de reciprocidad. Supongamos que el gradiente de temperaturas es cero. Entonces, el calor es únicamente de Peltier, $J=MV$, y usando la relación de Kelvin $M= T L$ y la definición de la termopotencia, se deduce inmediatamente que $S=Q/eT$, donde $Q$ es la energía promedio en forma de calor y $e$ es la unidad de carga elemental. Ahora bien, la fórmula de la entropía de Clausius es $Q/T$, por lo que podemos entonces interpretar el coeficiente de Seebeck como una forma de cuantificar la entropía que llevan los portadores por unidad de carga. Esta justificación que damos es cualitativa; con todo, es posible desarrollar un argumento riguroso. Esta propiedad de la termopotencia la hace sumamente interesante en estudios que intentan detectar estados exóticos en materiales cuánticos,
puesto que la entropía de estos estados se aleja de los valores que muestran los conductores convencionales. 

\section{Desaprender la termoelectricidad}

Vamos ahora a profundizar un poco más en las diferencias
entre los efectos termoeléctricos que se observan en
conductores macroscópicos y los que están presentes
en la nanoescala. Para este fin, consideremos el caso de dos puntos cuánticos acoplados capacitivamente. La ocupación media de cada uno de ellos puede fluctuar entre 0 y 1 electrón por estar acoplados a terminales con una temperatura $T$. Cuando uno de ellos está conectado a dos contactos metálicos, como en la figura 2, puede servir como el conductor donde se genera potencia. Si el otro está conectado a un único terminal, no contribuirá al transporte de carga; pero si dicho terminal está a una temperatura $T_h>T$, funcionará como fuente de calor. El acoplamiento del sistema con la fuente de calor estará pues mediada por la interacción repulsiva entre los electrones en puntos cuánticos distintos. A diferencia de los sistemas de dos terminales, donde el calor es transportado por los mismos electrones que generan la potencia, esta nueva configuración multiterminal permite separar la corriente eléctrica de la corriente de calor, de forma que puedan ser manipuladas independientemente~\cite{hotspots,holger,roche}.

\begin{figure}[t]
\includegraphics[width=0.5\linewidth]{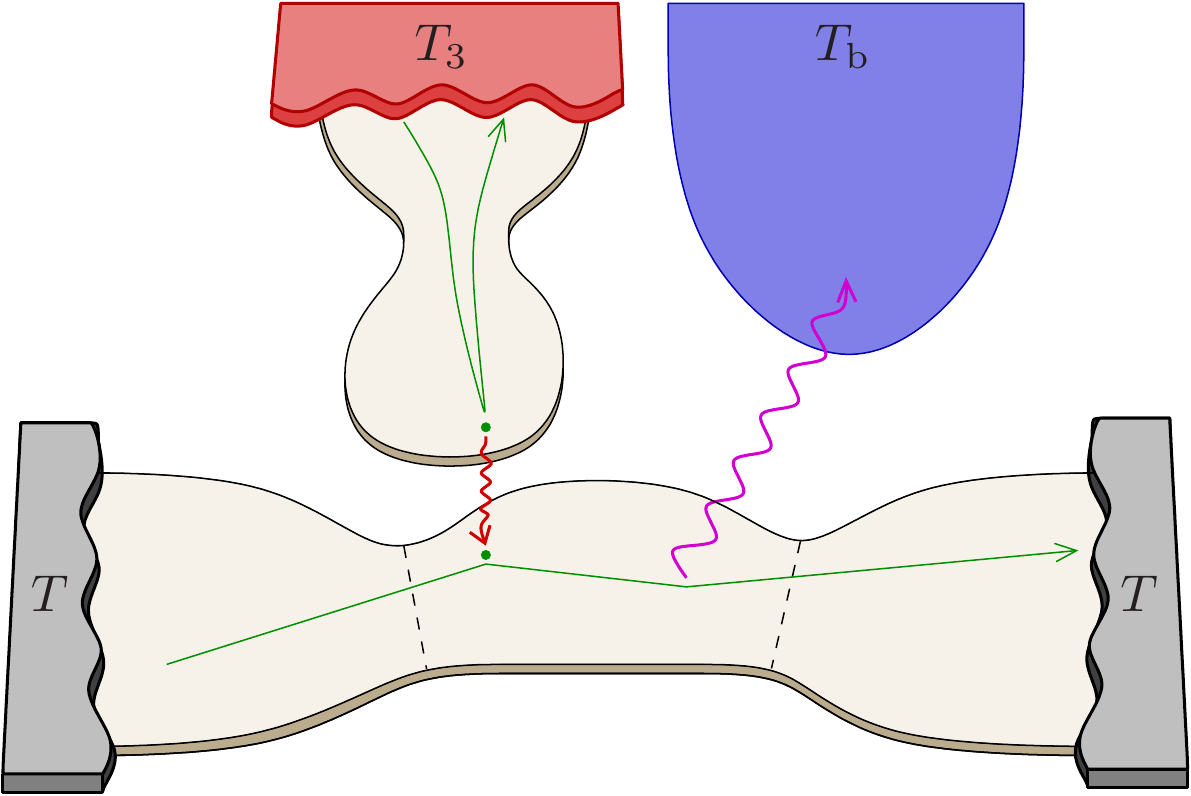}
\caption{\label{neq}
Configuraciones multiterminales permiten separar la fuente de calor del circuito termoeléctrico.
En el ejemplo del esquema, los electrones absorben calor de un terminal a temperatura $T_3$ mediante la interacción coulombiana entre electrones y disipan energía en un entorno bosónico a temperatura $T_{\rm b}$ mediante la emisión de fotones.}
\end{figure}

La combinación de varios de estos baños, como se muestra en la figura~\ref{neq}, permite explorar la física de estados fuera del equilibrio donde, por ejemplo, un electrón al propagarse interaccione a la vez y sin termalizar con una fuente de calor y con  un baño frío. De esta forma es posible que un sistema, incluso en casos que mantienen la simetría electrón-hueco, genere potencia\ldots\ ¡sin absorber calor!~\cite{libro}. 

Por otra parte, los sistemas de tres terminales adquieren nuevas propiedades cuando se les aplica un campo magnético. Esto hace que se rompa la simetría de inversión temporal, la propagación de los electrones adquiere quiralidad, afectando a las relaciones de reciprocidad de Onsager explicadas antes. En consecuencia, se deshace la relación entre los coeficientes de Seebeck y de Peltier, por lo que estos sistemas podrían comportarse indistintamente como máquinas de calor o como refrigeradores según se invierta la dirección del campo magnético~\cite{libro}.

En definitiva, indagar en las escalas nanoscópicas permite desentrañar los procesos responsables del efecto termoeléctrico y demontarlos como se desmonta un reloj. Si las piezas del reloj macroscópico son la disipación, la ruptura de la simetría electrón-hueco y la relación entre Seebeck y Peltier, al volver a montar el reloj mesoscópico puede que nos sobren todas las piezas.

\section{Aplicaciones}

Además de los argumentos fundamentales que apoyan el interés de los conductores cuánticos termoeléctricos, no hay que perder de vista sus aplicaciones prácticas. En ge\-ne\-ral,
se buscan sistemas que conviertan, por efecto Seebeck, el flujo de calor absorbido del entorno, $J$, en potencia útil, $P=IV$, donde la corriente generada $I$ es capaz de fluir en oposición a un voltaje $V$, y que esto ocurra con una eficiencia $\eta=P/J$ suficientemente alta. 

Si pensamos en el tamaño de los circuitos de cualquier aparato electrónico, y lo mucho que se calientan al funcionar, las posibilidades de aprovechar este calor residual son enormes. Por ejemplo, se pueden integrar circuitos que transformen el calor disipado a su alrededor para alimentar otras componentes del circuito y así disminuir su consumo energético y la necesidad de refrigerarlo. Por otra parte, se puede enfriar localmente un circuito mediante refrigeradores que utilicen el efecto Peltier. En el laboratorio, se ha llegado a enfriar contactos metálicos
termoeléctricamente
a temperaturas por debajo de los 50~mK~\cite{giazotto}.

En 1993, Mildred Dresselhaus se dio cuenta de que el espectro discreto que caracteriza a los sistemas nanoscópicos permitiría aumentar la eficiencia, lo cual se ha verificado posteriormente en diversas configuraciones~\cite{dresselhaus}. El motivo es que actúan como filtros de energía que maximizan la asimetría electrón-hueco. Además, si todos los electrones que atraviesan el sistema poseen la misma energía, las corrientes de carga y de calor son proporcionales. Esto hace que la eficiencia aumente linealmente con el voltaje hasta el punto en que $V$ cancela la corriente. Entonces, se alcanzaría el límite de Carnot donde  la conversión de calor es un proceso reversible y la entropía no aumenta. Por desgracia, esto también implica que no se produce potencia. En la práctica, se han conseguido eficiencias que llegan al 70\% de la de Carnot mediante un punto cuántico~\cite{josefsson}, generando una potencia finita aunque muy pequeña. En comparación, la eficiencia típica de un generador termoeléctrico macroscópico es del $15$\%. Estas altas eficiencias cuánticas se logran gracias a la alta repulsión coulombiana dentro del punto, que hace que los electrones solo puedan atravesar el sistema de uno en uno, como en un besamanos (efecto denominado bloqueo de Coulomb). Con todo, en la mayoría de los casos lo que interesa no es alcanzar la máxima eficiencia posible, sino, dada una $P$ necesaria para una operación concreta, maximizar la eficiencia correspondiente. Para ello, contamos con una gran riqueza de fenómenos presentes en los conductores cuánticos (resonancias, interferencias, interacciones) que pueden combinarse y que son además altamente controlables mediante campos externos~\cite{casati}. La termoelectricidad cuántica ya ha demostrado que es capaz de proporcionar dispositivos únicos, así que muy probablemente nos sorprenda con nuevas aplicaciones en el futuro.

\section{Agradecimientos}
D.\ S. agradece a R.\ Toral y R.\ Huecas por sus comentarios críticos del artículo.

\end{document}